\documentclass[12pt]{article}
\usepackage[utf8]{inputenc}
\usepackage{textcomp} %non-italicised µ
\usepackage{graphicx}
\usepackage{ulem}
\usepackage{amssymb}
\usepackage{amsmath}

\usepackage{color}
\usepackage[a4paper, portrait, margin=1.5cm]{geometry}

\begin{document}

\title{Generating phase singularities using surface exciton polaritons in an organic natural hyperbolic material}
%\author{Philip A. Thomas, William P. Wardley and William L. Barnes}
\author{Philip A. Thomas\footnote{Email: p.thomas2@exeter.ac.uk}, William P. Wardley \\ and William L. Barnes \\ \\ \small{Department of Physics and Astronomy, University of Exeter,} \\ \small{Exeter, EX4 4QL United Kingdom}}

\date{}

\maketitle

\begin{abstract}
    \noindent
    Surface polaritons (SPs) are electromagnetic waves bound to a surface through their interaction with charge carriers in the surface material.
    Hyperbolic SPs can be supported by optically anisotropic materials where the in-plane and out-of-plane permittivies have opposite signs.
    Here we report what we believe to be the first experimental study of hyperbolic surface exciton polaritons (HSEPs).
    We study the intensity and phase response of HSEPs in the J-aggregate TDBC (a type-II natural hyperbolic material).
    HSEPs can be used to generate phase singularities; the behaviour of these phase singularities is a consequence of the hyperbolic nature of TDBC.
    The combined intensity and phase response of non-hyperbolic and hyperbolic SPs suggests that they are topologically distinct.
    We predict analogous effects for hyperbolic surface phonon polaritons in hexagonal boron nitride.
    Our work suggests that organic materials can provide a new platform for the exploration of hyperbolic surface polaritonics at visible frequencies.
    
\end{abstract}

%%%Introduction
\newpage

\noindent
A key goal of nanophotonics is to rationally control the behaviour of light with sub-wavelength structures~\cite{novotny2012principles, basov2016polaritons}.
One of the most widely-utilised phenomena in the nanophotonic toolbox is the surface polariton (SP): the coupling of light and matter at an interface between two materials that occurs when the real components of their permittivities, Re($\epsilon$), have opposite signs~\cite{maier2007plasmonics}.
Surface plasmon polaritons (SPPs), in which light couples to free (conduction) electrons, are the most widely studied SP and have been explored in multiple spectral bands, including the infrared~\cite{berini2009long, grigorenko2012graphene}, visible~\cite{maier2007plasmonics, sonnichsen2002drastic} and ultraviolet~\cite{knight2014aluminum, wardley2019improving} regions.
Many other SP modes exist~\cite{basov2020polariton}.
Surface phonon polaritons (SPhPs), which rely on phonon modes to create spectral regions with Re($\epsilon$) $<0$, have been studied in the infrared and terahertz regions~\cite{caldwell2014sub, caldwell2015low}.
Surface exciton polaritons (SEPs) can occur in materials where spectral regions with Re($\epsilon$) $<0$ arise due to excitonic resonances~\cite{philpott1978exciton, philpott1979new}.
Phononic and excitonic resonances can both be modelled using Lorentz oscillators~\cite{caldwell2015low, philpott1979new}, a consequence of their shared physics, in spite of their different origins.

The investigation of all-organic nanophotonics has motived the study of SPPs in conducting polymer nanostructures, which - while currently restricted to the infrared spectral region - show promise because they can be modulated between plasmonic and dielectric behaviour~\cite{chen2023dynamic}.
SEPs provide a pathway to all-organic nanophotonics at visible frequencies since they can be generated using Frenkel excitons in organic dyes~\cite{gentile2014optical}.
Although the fundamental physics of SEPs has been well-understood for some decades~\cite{philpott1978exciton, philpott1979new}, experimental work has largely been limited to the study of propagating SEP modes in planar films~\cite{gu2013quest, gentile2014optical, nunez2016excitonic, takatori2017surface}.
Studies of excitonic nanostructures have been predominantly theoretical~\cite{gentile2016localized, humphrey2016excitonic, gentile2017hybridised, dutta2023effect, dutta2024weak} because of the challenges associated with creating organic nanostructures with sufficiently high dye concentrations to support localised SEP modes.
However, a recent report of the successful creation of excitonic nanostructures shows that these challenges can be overcome and that the field might be on the cusp of a transformation~\cite{kang2022organic}.

Optically anisotropic materials are said to be hyperbolic if their in-plane and out-of-plane permittivities have opposite signs~\cite{poddubny2013hyperbolic}.
The unusual dispersion relations of hyperbolic materials means they can be used to design unusual optical properties such as hyperlensing~\cite{liu2007far} and negative refraction~\cite{hoffman2007negative}.
While it is possible to design artificial hyperbolic metamaterials~\cite{poddubny2013hyperbolic}, the potential of natural hyperbolic materials has also been explored, with the hyperbolic SPhPs (HSPhPs) in hexagonal boron nitride (hBN) perhaps being the most well-known example~\cite{caldwell2014sub, caldwell2015low}.
However, very few natural hyperbolic materials have been reported at visible wavelengths~\cite{korzeb2015compendium}.

In this Letter, we report what we believe to be the first experimental study of hyperbolic surface exciton polaritons (HSEPs).
We show that the J-aggregate TDBC is highly anisotropic and can be classified as a type-II natural hyperbolic material with giant optical anisotropy.
We study prism-coupled HSEPs using spectroscopic ellipsometry, which allows us to simultaneously characterise the intensity and phase response of HSEPs~\cite{carini2025surface}.
We show that HSEPs can be used to completely suppress the reflection of light, resulting in the generation of a phase singularity, and that the distinctive phase response we observe is a consequence of the hyperbolic nature of the SEPs we studied.
We conduct numerical studies to consider the implications of our findings for analogous experiments using HSPhPs in the infrared.
Our work identifies a means of studying room-temperature natural hyperbolic SPs at visible frequencies.

%%%Results

\begin{figure}[!t]
    \centering
    \includegraphics[width=0.50\linewidth]{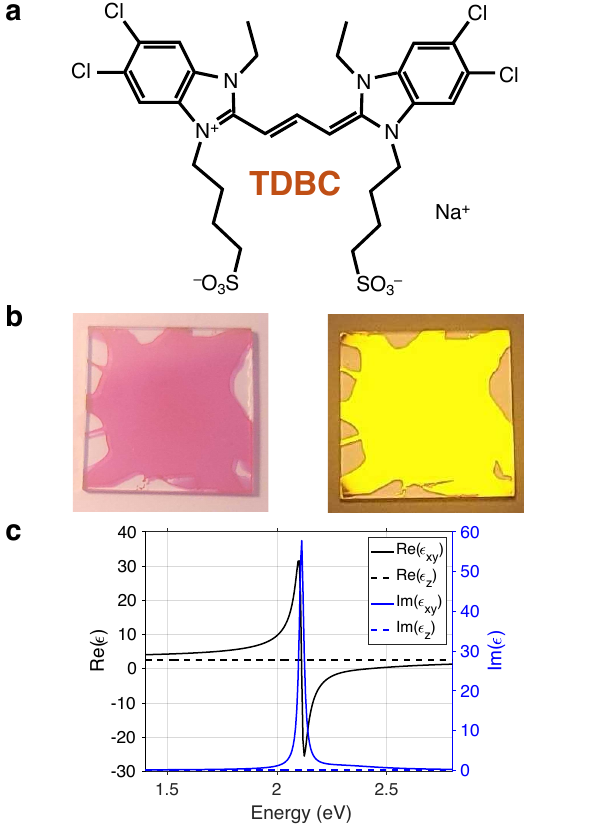}
    \caption{(a) Chemical structure of TDBC.
    (b) Photograph of neat TDBC film under ambient illumination (left) and top-down illumination, reflecting light into the camera (right); the substrate is 25 mm $\times$ 25 mm.
    (c) Optical constants of TDBC, expressed in terms of in-plane and out-of-plane permittivities ($\epsilon_\text{xy}$ and $\epsilon_\text{z}$, respectively).
    }
    \label{fig:eps}
\end{figure}

%%%Why use TDBC?

%%\textcolor{gray}{\sout{
%%\textcolor{blue}{

We studied the J-aggregate TDBC (5,6-dichloro-2-[[5,6-dichloro-1-ethyl-3-(4-sulphobutyl)-benzimidazol-2-ylidene]-propenyl]-1-ethyl-3-(4-sulphobutyl)-benzimidazolium hydroxide, sodium salt, inner salt).
This dye (chemical structure in Figure \ref{fig:eps}a) possesses one sharp excitonic resonance in the visible spectrum, which has made it an appealing candidate in molecular strong coupling experiments~\cite{dintinger2005strong, torma2014strong}.
In high enough concentrations, TDBC films can possess a band of negative permittivity that is sufficiently negative and broad to make SEP experiments viable~\cite{gentile2014optical, takatori2017surface, kang2022organic}.
TDBC is part of a wider family of J-aggregates that can be used to create negative-permittivity films at wavelengths across the visible and near-infrared spectrum~\cite{holder2023bio}: while our experiments focus solely on TDBC, we expect our findings may be generalised to many dyes in this class of J-aggregates~\cite{wurthner2011j}.
This would allow one to use natural hyperbolic materials across a wide range of wavelengths, overcoming the main limitation of natural hyperbolic materials~\cite{takamaya2019optics}.
We discuss other candidate materials in Supplementary Section S1.1.

%%%What do we learn from the optical constants of TDBC?

Figure \ref{fig:eps}b shows photographs of one of the TDBC films made in our experiments.
The optical constants of this film, determined using spectroscopic ellipsometry, are shown in Figure \ref{fig:eps}c (see Supplementary Section S1.2 for fabrication and characterisation details).
The high concentration of our film allows us to reach a minimum Re($\epsilon$) $= -25$, with Re($\epsilon$) $< -2$ (the condition for propagating SP modes~\cite{maier2007plasmonics}) in the spectral range 2.11--2.26 eV.
These numbers are consistent with the optical constants for pure TDBC films reported by Kang \textit{et al.}~\cite{kang2022organic} and suggest that we have a broad enough spectral window to observe dispersive SEP modes in a prism coupling experiment.
We studied films of thickness between 30 nm and 200 nm: in all cases, we found that it was only possible to achieve a good fit to ellipsometric data using an anisotropic model.
J-aggregate films are structurally anisotropic, with aggregates aligning in the plane of the substrate, which should naturally give rise to optical anisotropy~\cite{roodenko2013anisotropic}.
While our reported experiments focus on neat TDBC films, we found that this anisotropy persists for TDBC dissolved in PVA (see Supplementary Section S1.3).
In spite of this, we are aware of only a small number of experimental works that explicitly acknowledge the optical anisotropy of TDBC~\cite{hayashi2012plasmonic, kang2022organic}, with most experiments modelling TDBC as an optically isotropic material.
Additional analysis in Supplementary Section S1.4 provides further support for these optical constants, suggesting that it is possible to create TDBC films with near-perfect aggregation.
Since the out-of-plane permittivity is always positive, we can classify TDBC films as a Type-II natural hyperbolic material in the region where the in-plane permittivity is negative (2.11--2.43 eV)~\cite{caldwell2014sub}.
Neat TDBC films also fulfil the criterion for giant optical anisotropy; see Supplementary Section S3.

%%%Prism coupling experiment using ellipsometry
%%\textcolor{gray}{\sout{
%%\textcolor{blue}{

\begin{figure}[!b]
    \centering
    \includegraphics[width=0.5\linewidth]{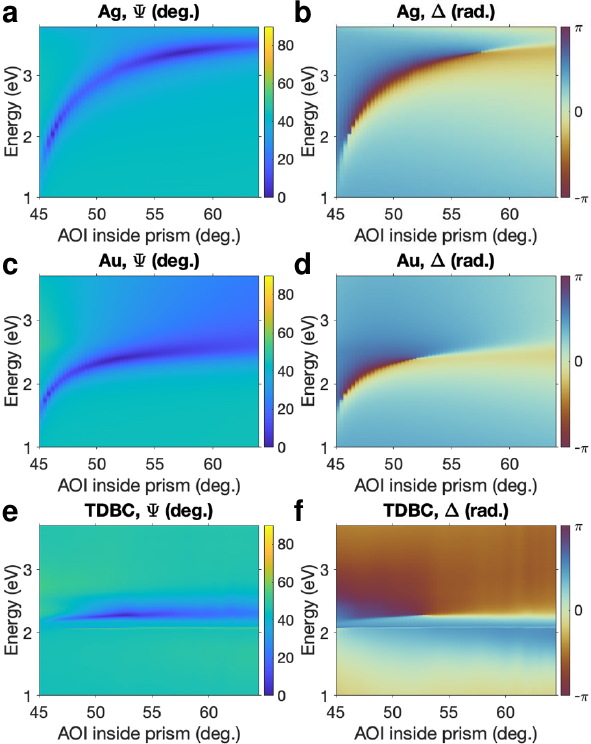}
    \caption{Results from ellipsometric prism coupling experiments in which SPPs are observed in Ag (panels a,b) and Au (panels c,d) and an SEP mode is observed in TDBC (panels e,f).
    Dispersion plots are constructed in panels a, c, e using $\Psi$ (intensity) data and using $\Delta$ (phase) data in panels b, d, f.
    The TDBC dispersions are replotted in Figure \ref{fig:uniaxial} with a zoomed energy axis.}
    \label{fig:prism}
\end{figure}

To characterise the HSEPs in our TDBC films, we performed a prism coupling experiment using the attenuated total reflection technique in the Kretschmann-Raether prism coupling geometry~\cite{kretschmann1968radiative}.
(See Supplementary Section S2.1 for further details.)
We characterised the HSEPs using spectroscopic ellipsometry~\cite{tompkins2005ellipsometry}, which measures $\rho$, the complex reflectance ratio:
\begin{align*}
    \rho = \frac{r_p}{r_s} = \tan(\Psi)e^{i\Delta},
\end{align*}
\noindent where $r_p$ and $r_s$ are the Fresnel reflection coefficients for p- and s-polarised light, respectively. $\tan(\Psi)$ is the ratio of the magnitudes of the amplitudes of p- and s-polarised light: $\Psi = 0^{\circ}$ when $r_p = 0$ and $\Psi = 90^{\circ}$ when $r_s = 0$.
For propagating surface polariton modes, which are only excited by p-polarised light, we therefore expect to observe minima in $\Psi$.
$\Delta$ is the difference in phase between p- and s-polarised light.
Further details concerning ellipsometry are provided in Supplementary Section S2.1.

In Figure \ref{fig:prism} we plot the results from our prism coupling experiments: the top row (2a-b) shows results from prism coupling to a Ag film (thickness 50 nm), the middle row (2c-d) shows results using a Au thin film (thickness 40 nm) and the bottom row (2e-f) shows results using a TDBC thin film (thickness 90 nm).
Our ellipsometer has an angle range of $45^\circ$-$75^\circ$; we have plotted our results using the incident angle of light from within the prism ($45^\circ$-$65^\circ$).
The left column (Figs. \ref{fig:prism}a,c,e) shows dispersion plots constructed using $\Psi$; the dispersion plots in the right column (Figs. \ref{fig:prism}b,d,f) are constructed using $\Delta$.
A detailed analysis of these results, including a comparison of quality factors and an analysis of $\rho$ plots, can be found in Supplementary Section S2.
Here, we highlight three points.
First, the observed SPP modes in the Ag and Au $\Psi$ plots (Figs. \ref{fig:prism}a and \ref{fig:prism}c) closely match those calculated using standard literature values~\cite{palik1985handbook}.
Second, the TDBC $\Psi$ plot (\ref{fig:prism}e) differs from the Ag and Au plots in two ways: there is an essentially non-dispersive feature at 2.1 eV, which corresponds to the maximum in Re($\epsilon$) for TDBC, and the dispersive feature at 2.2-2.5 eV (re-plotted for this narrow spectral region in Figure \ref{fig:uniaxial}) is the HSEP mode.
Third, we notice that all $\Delta$ plots contain phase singularities, points at which reflection is completely suppressed~\cite{berry2000making, nye1974dislocations}.
The kind of phase singularity observed here - observed not in real space, but in Fourier space - has been investigated for applications in sensing and modulation~\cite{ni2021multidimensional, kabashin2023label}.
The Ag and Au $\Delta$ plots (\ref{fig:prism}b,d) both show two phase singularities generated by the SPP modes.
In contrast, there is only one phase singularity associated with the HSEP in Fig. \ref{fig:prism}f.

%%%The role of anisotropy

\begin{figure}[!b]
    \centering
    \includegraphics[width=0.5\linewidth]{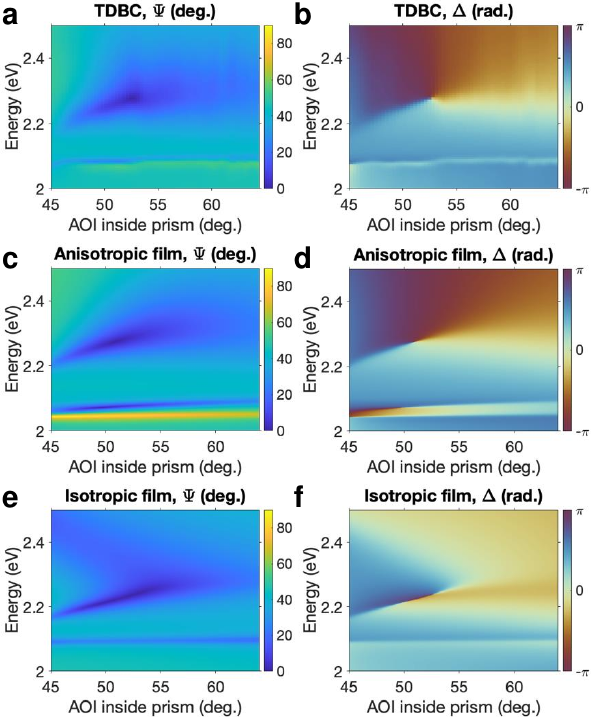}
    \caption{The effect of anisotropy on the SEP dispersion.
    (a,b) Experimental $\Psi$ and $\Delta$ dispersions from TDBC prism coupling experiments (previously plotted in Figure \ref{fig:prism}e,f).
    (c,d) Calculated $\Psi$ and $\Delta$ dispersions using a uniaxial model of TDBC (optical constants in Figure \ref{fig:eps}c), with TDBC film thickness set to 90 nm.
    (e,f) Calculated $\Psi$ and $\Delta$ dispersions using an isotropic model of TDBC ($\epsilon_{\text{x,y}}$ in Figure \ref{fig:eps}c), with TDBC film thickness set to 44 nm.
    }
    \label{fig:uniaxial}
\end{figure}

To understand the difference between SEP and HSEP behaviour, in Figure \ref{fig:uniaxial}a-b we re-plot the $\Psi$ and $\Delta$ dispersions in Figure \ref{fig:prism}e-f, rescaling to focus on the spectral region in which TDBC's HSEP mode is observed.
We attempt to reproduce these plots using standard transfer matrix calculations in Figure \ref{fig:uniaxial}c-f.
In Fig. \ref{fig:prism}c-d, we model TDBC with the uniaxial optical constants in Figure \ref{fig:eps}c and find good agreement with experiments: the HSEP mode in $\Psi$ and its associated phase singularity in $\Delta$ are both accurately reproduced.
The two modes between 2.0 and 2.1 eV (quasi-normal modes associated with the narrow region of giant optical anisotropy, see Supplementary Section S3) are also reproduced, although they appear substantially weaker in experiments (likely due to roughness of the films).
In Fig. \ref{fig:prism}e,f, we model TDBC with isotropic optical constants (using the in-plane values from Fig. \ref{fig:eps}c).
We can reproduce the $\Psi$ response with a SEP by reducing the TDBC thickness from 90 nm to 44 nm; however, this model produces two phase singularities in $\Delta$, more closely resembling the Ag and Au SPP dispersions (Figures \ref{fig:prism}b, d) than the SEP dispersion, which has just one phase singularity.
Furthermore, the isotropic model predicts back-bending of the SEP mode at 2.3-2.5 eV which is absent in the experimental data and uniaxial calculations.
The isotropic model predicts one TM mode associated with giant optical anisotropy (a minimum in $\Psi$), while the experimental and uniaxial plots both show and additional TE mode (a maximum in $\Psi$).
These calculations show that, although there are clear differences between SEPs and HSEPs in intensity measurements, the most striking differences occur in phase measurements: SEPs produce two phase singularities, while HSEPs produce only one.
This suppression of reflection can be considered a form of critical coupling~\cite{tischler2007critically}.
The conditions for zero reflection within a structure can be rationally designed by carefully tuning its material or geometric properties~\cite{thomas2022all, cusworth2023topological}.
Hyperbolic materials are inherently more complex than their isotropic counterparts, meaning that the conditions for critical coupling are stricter (especially when a material has no out-of-plane absorption).
Therefore, it is unsurprising that HSEPs produce fewer phase singularities than isotropic SEPs.
In Supplementary Section S4 we analyse ellipsometry for TDBC films with a wide range of thicknesses (30 nm -- 200 nm) in the non-prism geometry (i.e. air superstrate/TDBC/glass substrate).
While anisotropy radically alters the dispersion of SEP/HSEP modes, these results suggest that the effect of anisotropy on other photonic modes in these structures are much more subtle.

%%%Generalising: hBN

\begin{figure}[!b]
    \centering
    \includegraphics[width=0.5\linewidth]{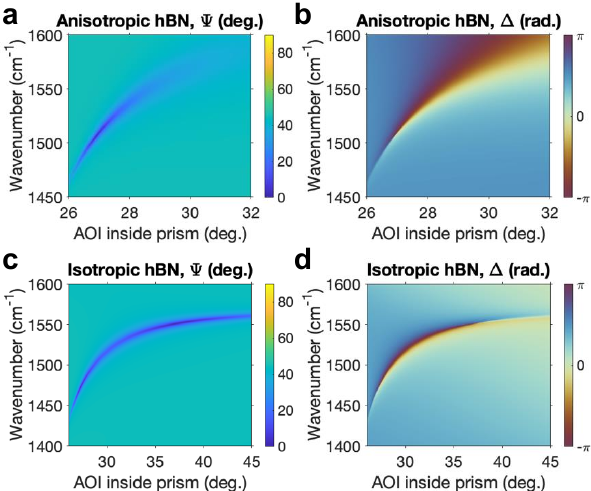}
    \caption{The effect of anisotropy on prism coupling to HSPhPs in hBN.
    (a,b) Calculated $\Psi$ and $\Delta$ dispersions using a uniaxial model of hBN, with hBN film thickness set to 1.5 µm.
    (c,d) Calculated $\Psi$ and $\Delta$ dispersions using an isotropic model of hBN (setting $\epsilon_{\text{z}}=\epsilon_{\text{x,y}}$), with hBN film thickness set to 800 nm.
    }
    \label{fig:hbn}
\end{figure}

While our experiments have focused on HSEPs, the universal nature of the Lorentz oscillator model means our findings should generalise to other hyperbolic SPs.
In Fig. \ref{fig:hbn} we plot the calculated dispersion of prism-coupled HSPhPs in hBN.
We have performed our calculations for the spectral region for which hBN is a type-II natural hyperbolic material~\cite{caldwell2014sub}; see Supplementary Section S5 for further details.
In Figs. \ref{fig:hbn}a-b we plot $\Psi$ and $\Delta$ dispersions for uniaxial hBN and observe a HSPhP dispersion in $\Psi$ with a single associated phase singularity in $\Delta$, closely resembling the experimental and calculated HSEP dispersions in Fig. \ref{fig:uniaxial}a-d.
When we model hBN as an optically isotropic material in Figs. \ref{fig:hbn}c-d, the non-hyperbolic SPhP dispersion much more closely resembles the Ag SPP dispersion in Figs. \ref{fig:prism}a-b (both in terms of the dispersion in $\Psi$ and the observation of two phase singularities in $\Delta$).
The strong parallels between our HSEP and HSPhP results suggest that HSPhPs might be used to generate phase singularities in the infrared~\cite{carini2025surface}.
They also suggest that HSEPs in J-aggregates such as TDBC might provide a viable platform to study hyperbolic SPs at visible wavelengths.

In Figure \ref{fig:rhoSP} we combine $\Psi$ and $\Delta$ to plot the values of $\rho$ corresponding to the hyperbolic and non-hyperbolic SP modes calculated in Figs. \ref{fig:uniaxial}c-f and \ref{fig:hbn} (see Supplementary Section S6 for details of extraction method).
The isotropic SEP and SPhP modes each generate two phase singularities, meaning that they cross $\rho=0$ twice, creating a loop in $\rho$.
(The SEP loop is smaller because the two phase singularities are much closer together in Fourier space, meaning the SEP amplitude stays much lower between the two $\rho=0$ points.)
Meanwhile, since the HSEP and HSPhP modes each only support one phase singularity, they only cross $\rho=0$ once, eliminating this loop.
The clear difference between the topologies of hyperbolic and non-hyperbolic SPs (a consequence of their different phase responses) perhaps hints at the more profound differences between hyperbolic and non-hyperbolic SPs~\cite{thomas2020new}.
While beyond the scope of this Letter, a similar approach could be used to study the response of metasurfaces, which are frequently inherently anisotropic~\cite{meinzer2014plasmonic}.

\begin{figure}[!t]
    \centering
    \includegraphics[width=0.5\linewidth]{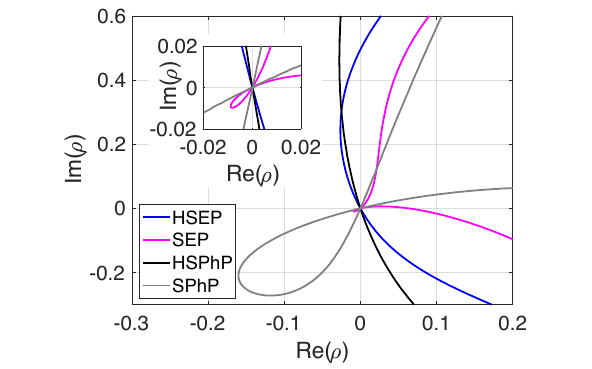}
    \caption{The combined amplitude and phase responses of HSEP (blue line), SEP (magenta line), HSPhP (black line) and SPhP (grey line) modes.
    Inset: zooming in on $\rho=0$ confirms that the plotted SEP and SPhP have the same topology.
    }
    \label{fig:rhoSP}
\end{figure}

%%%Conclusion

In conclusion, we have introduced the idea of hyperbolic surface exciton polaritons and shown that they can be supported by J-aggregates, an overlooked example of a natural hyperbolic material.
We have shown for the first time that surface exciton polaritons can be used to generate phase singularities; our calculations suggest that a similar phase response can be generated using surface phonon polaritons.
Our combined intensity and phase analysis has revealed that hyperbolic and non-hyperbolic surface polaritons are topologically distinct.
Our work shows that organic natural hyperbolic materials provide a viable platform for the room-temperature study of hyperbolic surface polaritons in the visible spectrum.
\\

\newpage
The data that support the findings of this study are openly available in University of Exeter data repository at XXXX. \\

%\section*{Acknowledgements}
%\begin{acknowledgments}
Funded in part by the European Union (Project SCOLED, Grant Agreement Number 101098813). Views and opinions expressed are however those of the authors only and do not necessarily reflect those of the European Union or the European Innovation Council and SMEs Executive Agency (EISMEA). Neither the European Union nor the granting authority can be held responsible for them. For the purpose of open access, the authors had applied a Creative Commons Attribution (CC BY) licence to any Author Accepted Manuscript version arising.
%\end{acknowledgments}

%\title{Supplementary Information for:\\ Generating phase singularities using surface exciton polaritons in an organic natural hyperbolic material}

%\author{Philip A. Thomas, William P. Wardley \& William L. Barnes}
%\date{}

\newpage
\begin{center}
    %\\ \vspace{5cm}
    {\LARGE  Supplementary Information for:}
    \\ \vspace{0.5cm}
    {\LARGE Generating phase singularities using surface exciton polaritons in an organic natural hyperbolic material}
    \\ \vspace{0.5cm}
    Philip A. Thomas, William P. Wardley and William L. Barnes,
\\
\small{Department of Physics and Astronomy, University of Exeter,} \\ \small{Exeter, EX4 4QL, United Kingdom} \\
\end{center}

\maketitle

\tableofcontents

\newpage
\addcontentsline{toc}{section}{S1. Optical constants of TDBC films}
\section*{S1. Optical constants of TDBC films}

\addcontentsline{toc}{subsection}{S1.1 Comparison of TDBC to other candidate materials}
\subsection*{S1.1 Comparison of TDBC to other candidate materials}

When designing a SEP experiment, one needs to select a dye which permits sufficiently negative Re($\epsilon$).
This can be achieved using a high concentration of a dye molecule with a large dipole moment (equivalent to a high oscillator strength $f$) that yields a spectrally narrow resonance (low damping $\gamma$).
While it might be tempting to take a well-known, highly absorbing dye and dramatically increase its concentration until negative permittivity is achieved, such an approach fails to account for the increase in $\gamma$ that accompanies very large increases in $f$~\cite{dutta2023effectsupp}.
Dutta \textit{et al.}~\cite{dutta2022modelingsupp} illustrated this with films of rhodamine dye, finding that - in spite of its large oscillator strength - at best it could only marginally give Re($\epsilon$) $<0$.

We are only aware of two other reported organic natural hyperbolic materials: squaraine (for which Re($\epsilon$) $<-2$ for an impractically narrow spectral window of 34 meV~\cite{kim2021naturalsupp}) and quinoidal oligothiophene derivatives (for which Re($\epsilon$) never drops below $-1$~\cite{lee2019organicsupp}), neither of which are promising candidates for surface exciton polaritonics.
It has also been suggested that transition metal dichalcogenides such as WS$_2$ might support hyperbolic surface exciton polaritons; however, the predicted negative permittivity window is below 5 meV and requires cryogenic temperatures~\cite{epstein2020highlysupp}.

\addcontentsline{toc}{subsection}{S1.2 Fabrication and optical constants of neat TDBC films}
\subsection*{S1.2 Fabrication and optical constants of neat TDBC films}

TDBC was purchased from FEW Chemicals.
To maximise the concentration of TDBC in our films, we fabricated pure TDBC films by dissolving TDBC in deionised water (concentration $3\%$wt TDBC in deionised water, sonicated for 25 minutes then stirred on a magnetic stirrer plate until homogenously coloured, at least 2 hours) and used spin-coating to form a thin film on a glass substrate (spin speeds 3 krpm -- 7 krpm).

Frequently the permittivity of a dye-doped film is modelled with the Lorentz oscillator model~\cite{fox2010opticalsupp}:
\begin{align*}
    \epsilon(\omega) = \epsilon_\infty + \sum_1^j\frac{\omega_j^2 f_j}{\omega_j^2 - \omega^2 - i\omega\gamma_j}.
\end{align*}
Each absorption peak is modelled by one or more oscillators with resonance frequency $\omega_j$, reduced oscillator strength $f_j$ and damping rate $\gamma_j$~\cite{gentile2014opticalsupp};
$\epsilon_\infty$ is the background permittivity of the film.
For a more extensive discussion on the applicability of the Lorentz oscillator model see \cite{dutta2022modelingsupp,stete2023opticalsupp}.

For the neat TDBC films discussed in the main manuscript, we used a uniaxial model for TDBC's optical constants.
For the in-plane component, $\epsilon_{\text{x,y}}$, we used  $\epsilon_\infty = 2.54$, $\omega_1 = 2.11$ eV, $\gamma_1 = 0.0250$ eV, $f_1 = 0.682$, $\omega_2 = 2.31$ eV, $\gamma_2 = 0.447$ eV, $f_2 = 0.197$.

For the out-of-plane component, we set $\epsilon_{z} = \epsilon_\infty = 2.54$.

\newpage
\addcontentsline{toc}{subsection}{S1.3 Optical constants of PVA-TDBC films}
\subsection*{S1.3 Optical constants of PVA-TDBC films}

\renewcommand{\thefigure}{S1}
\begin{figure}[!h]
    \centering
    \includegraphics[width=0.5\linewidth]{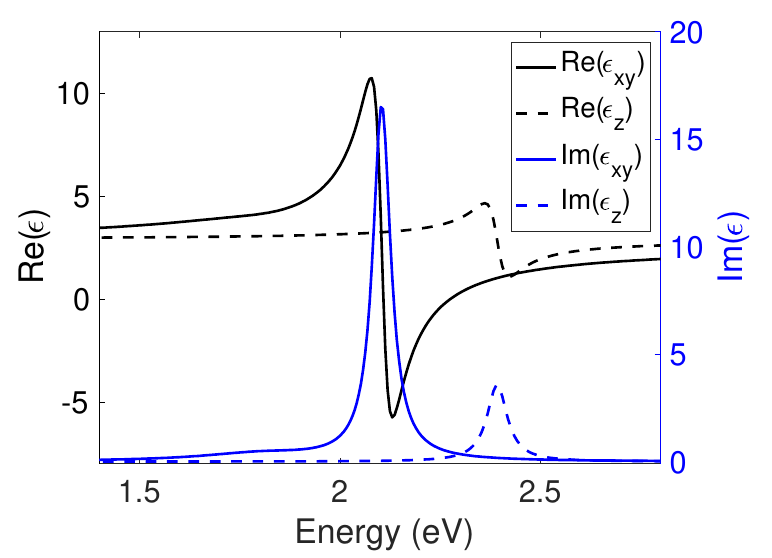}
    \caption{Optical constants of TDBC-doped PVA, determined using spectroscopic ellipsometry.}
    \label{fig:pva}
\end{figure}

\noindent
We performed spectroscopic ellipsometry on a TDBC-doped PVA film using the fabrication method outlined by Gentile \textit{et al.}~\cite{gentile2014opticalsupp}.
Our in-plane permittivity is almost identical to the permittivity determined by Gentile \textit{et al.}. We used $\epsilon_\infty = 2.55$, $\omega_1 = 2.11$ eV, $\gamma_1 = 0.0538$ eV, $f_1 = 0.422$, $\omega_2 = 1.81$ eV, $\gamma_2 = 0.372$ eV, $f_2 = 0.0739$.
For our out-of-plane permittivity, we used $\epsilon_\infty = 2.86$, $\omega_1 = 2.39$ eV, $\gamma_1 = 0.0609$ eV, $f_1 = 0.0910$.
The energy of the out-of-plane Lorentz oscillator corresponds to the absorption energy of unaggregated TDBC monomers~\cite{lidzey2015strongsupp}. These optical constants for TDBC in PVA - with an aggregated TDBC response in-plane and an unaggregated TDBC response out-of-plane - are consistent with the conclusions of the TCC film study performed by Roodenko \textit{et al.}~\cite{roodenko2013anisotropicsupp}.
Our results suggest that anisotropy is not restricted to neat TDBC films but might be present in many TDBC-doped films studied in the literature.
Importantly, these results show that, with a sufficiently high dye concentration, TDBC-doped PVA can still be classified as a type-II natural hyperbolic material with giant optical anisotropy.

\newpage
\addcontentsline{toc}{subsection}{S1.4 Adding Lorentz oscillators to the out-of-plane permittivity of TDBC}
\subsection*{S1.4 Adding Lorentz oscillators to the out-of-plane permittivity of TDBC}

\renewcommand{\thefigure}{S2}
\begin{figure}[!h]
    \centering
    \includegraphics[width=0.9\linewidth]{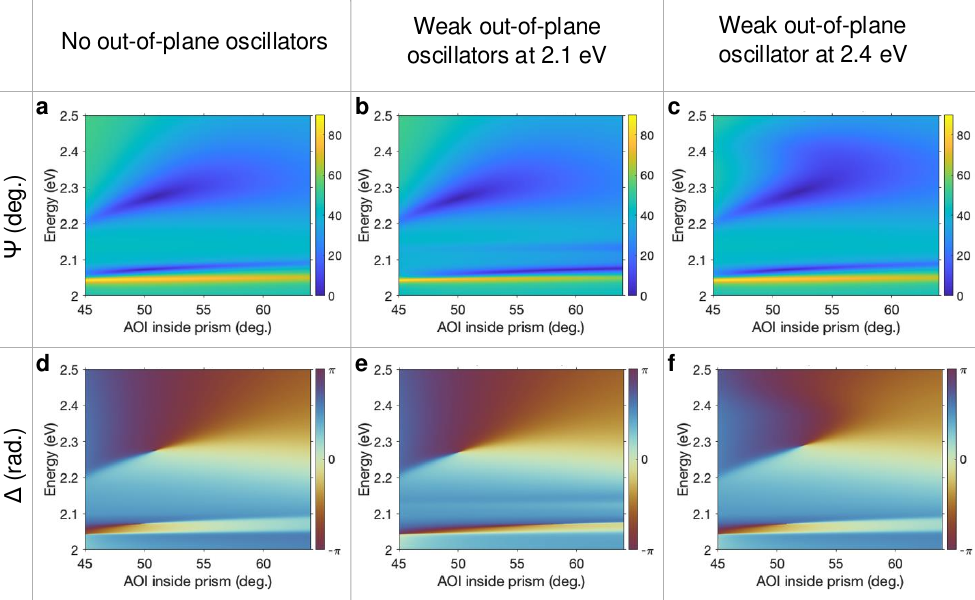}
    \caption{Calculated ellipsometric $\Psi$ (a-c) and $\Delta$ (d-f) prism coupling dispersions using different anisotropic models of TDBC.
    (a, d) Out-of-plane permittivity, $\epsilon_z$, has no Lorentz oscillators. (b, e) $\epsilon_z$ has two oscillators at 2.1 eV (like in-plane, but with much weaker oscillator strength). (c, f) $\epsilon_z$ has one oscillator at 2.4 eV.}
    \label{fig:oop}
\end{figure}

\noindent
Our optical constants for TDBC are anisotropic, with two Lorentz oscillator in-plane at 2.1 eV and 2.3 eV and and no oscillators out-of-plane.
This model gives an excellent match to experimental data both in the prism coupling geometry (Fig. 3 in the main manuscript) and in the air/TDBC/glass geometry (see Supplementary Figure \ref{fig:top}).
The calculated prism coupling dispersion using this model is reproduced in Figure \ref{fig:oop}a,d.
The lack of out-of-plane absorption at either 2.1 eV or 2.4 eV (the absorption energies of aggregated and unaggregated TDBC, respectively) imply that essentially all TDBC in the film is aggregated with dipoles lying in-plane, which is perhaps surprising.
Here we explore the effect of adding weak oscillators to the out-of-plane permittivity, $\epsilon_z$.

For panels b,e we use the same Lorentz oscillators out-of-plane as we do in-plane, but reduce the oscillator strength to $5\%$ of the in-plane values (i.e. for $\epsilon_z$, $f_1 = 0.0341$ and $f_2 = 0.00985$).
Even with these very small oscillator strengths, new spectral features appear around 2.1 eV which do not appear in the experimental data.
We therefore conclude that any out-of-plane aggregated TDBC has a negligible influence on the optical response of our TDBC film.
Our optical constants for TDBC are consistent with the anisotropic model of Kang \textit{et al.}~\cite{kang2022organicsupp}: taken together, our works suggest that it is possible to create TDBC films with near-perfect aggregation.

In panels c,f we introduce a new Lorentz oscillator at 2.4 eV ($f = 0.03$, $\gamma = 0.15$ eV, $\omega = 2.4$ eV), which roughly corresponds to the optical response expected from unaggregated TDBC monomers~\cite{lidzey2015strongsupp}.
This adds features to both $\Psi$ and $\Delta$ dispersions at 2.4 eV which do not correspond to any experimentally observed features.
The absence of any spectral features that correspond to unaggregated TDBC monomers (either in-plane or out-of-plane) suggests that only a negligible quantity of TDBC remains unaggregated.

\newpage
\addcontentsline{toc}{section}{S2. Further details of prism coupling surface polariton experiments}
\section*{S2. Further details of prism coupling surface polariton experiments}

\addcontentsline{toc}{subsection}{S2.1 The use of ellipsometry in prism coupling experiments}
\subsection*{S2.1 The use of ellipsometry in prism coupling experiments}

\renewcommand{\thefigure}{S3}
\begin{figure}[!h]
    \centering
    \includegraphics[width=0.8\linewidth]{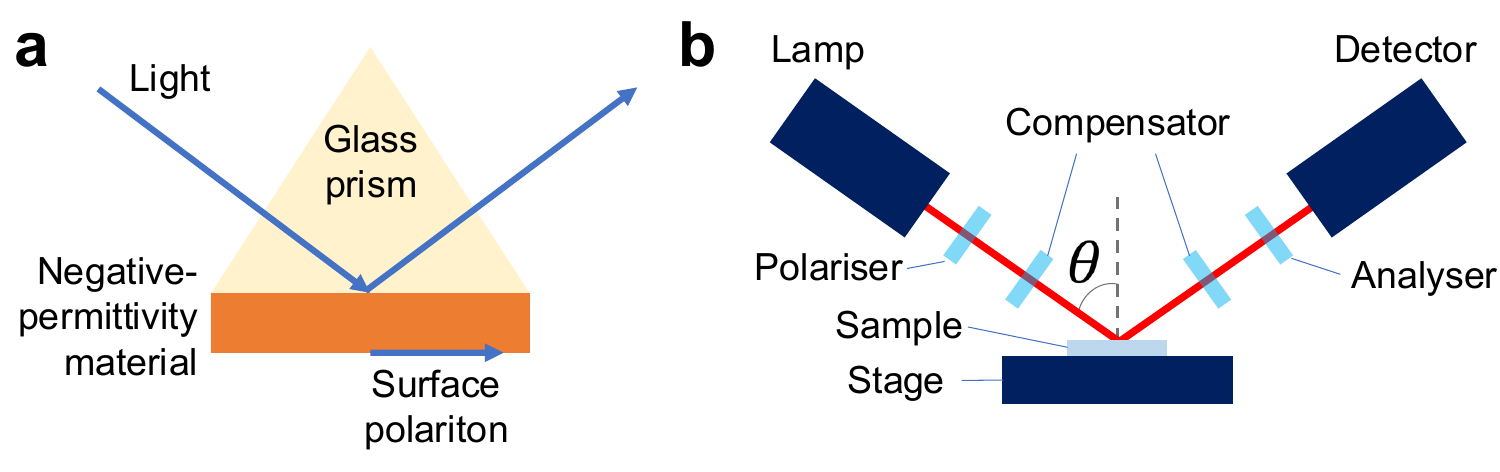}
    \caption{Experimental design. (a) Surface polariton prism coupling geometry. (b) Ellipsometer schematic.}
    \label{fig:prismellips}
\end{figure}

\noindent
Surface polaritons may enable sub-wavelength control because the light-matter interaction leads to the momentum of the surface polariton being greater than that of a free space photon~\cite{barnes2006surfacesupp}.
This momentum enhancement means that surface polaritons do not couple easily to radiating light: some form of momentum matching is required, examples include scattering~\cite{barnes2016particlesupp, sonnichsen2002drasticsupp}, prism coupling~\cite{lawrence1992criticalsupp, berini2009longsupp}, grating coupling~\cite{worthing2010couplingsupp, bryan2007diffractionsupp} and near-field dipole coupling~\cite{barnes1998fluorescencesupp, baieva2012strongsupp}.
Surface polaritons thus allow the usual diffraction limit of light to be overcome~\cite{koppens2011graphenesupp}.

In our experiments, we excited surface polaritons using the widely-used attenuated total reflection technique in the Kretschmann-Raether prism coupling geometry~\cite{kretschmann1968radiativesupp} (schematic in Fig. \ref{fig:prismellips}a).
We used a right-angle prism with refractive index $n=1.5$ and glycerol as the index-matching medium between the prism and glass substrate.
This excites surface polaritons at the interface between the negative-permittivity material and air.

Surface polariton experiments are commonly characterised using reflectivity measurements~\cite{kretschmann1968radiativesupp}; however, in our work, we follow a growing trend by characterising our SEP modes using spectroscopic ellipsometry (schematic in Fig. \ref{fig:prismellips}b) ~\cite{tompkins2005ellipsometrysupp}.
Spectroscopic ellipsometry measures $\rho$, the complex reflectance ratio:
\begin{align*}
    \rho = \frac{r_p}{r_s} = \tan(\Psi)e^{i\Delta}.
\end{align*}
$r_p$ and $r_s$ are the Fresnel reflection coefficients for p- and s-polarised light, respectively. $\tan(\Psi)$ is the ratio of the magnitudes of the amplitudes of p- and s-polarised light: $\Psi = 0^{\circ}$ when $r_p = 0$ and $\Psi = 90^{\circ}$ when $r_s = 0$.
For propagating surface polariton modes, which are only excited by p-polarised light, we expect to observe minima in $\Psi$.
$\Delta$ is the difference in phase between p- and s-polarised light.

Ellipsometry has recently found wider use in nano-optics because it allows for the simultaneous characterisation of the intensity and phase response of a sample~\cite{kravets2014graphenesupp, kravets2008extremelysupp, diest2013aluminumsupp, tsurimaki2018topologicalsupp, sreekanth2018biosensingsupp, thomas2020newsupp, ermolaev2022topologicalsupp, carini2025surfacesupp}.
It has advantages that make it appealing for prism-coupled surface polariton experiments.
There are no dispersive s-polarised electromagnetic modes in our surface polariton experiments, meaning that analysis of $\Psi$ spectra is no more complex than it is for reflectivity spectra.
For example, it is possible to use $\Psi$ spectra to compare the quality factors of different surface polariton modes as one would using reflectivity data~\cite{kravets2014graphenesupp}.
In fact, in SEP experiments one could argue that the analysis of $\Psi$ spectra is simpler than the analysis of reflectivity spectra.
In reflectivity measurements, one would expect to observe an increase in reflectivity (regardless of polarisation) in regions where there is a high impedance mismatch between TDBC and its surrounding materials.
Within this region of enhanced reflectivity, one would expect to observe a drop in reflectivity corresponding to the SEP.
In $\Psi$ measurements, normalising $r_p$ with respect to $r_s$ cancels out this enhancement of reflectivity, meaning that only the SEP mode remains.
Furthermore, the fact that ellipsometry measures the ratio of p- and s-polarised light makes it an ideal experimental technique for prism coupling experiments, where the normalisation of reflectivity measurements is otherwise challenging.
Indeed, our ellipsometric SEP dispersions plots are much lower-noise than those built using reflectivity measurements~\cite{gentile2014opticalsupp, gu2013questsupp, takatori2017surfacesupp}.

\addcontentsline{toc}{subsection}{S2.2 Further analysis of prism coupling surface polariton experiments}
\subsection*{S2.2 Further analysis of prism coupling surface polariton experiments}

\noindent
In the main manuscript we used surface plots to present the results of our ellipsometric prism coupling experiments.
These plots allow one to directly visualise the dispersions of surface polaritons.
Here we provide line plots of the same data. first of $\Psi$ and $\Delta$ as a function of incident angle (with fixed energy), then of $\Psi$ and $\Delta$ as a function of energy (with fixed incident angle).

\renewcommand{\thefigure}{S4}
\begin{figure}[!b]
    \centering
    \includegraphics[width=0.6\linewidth]{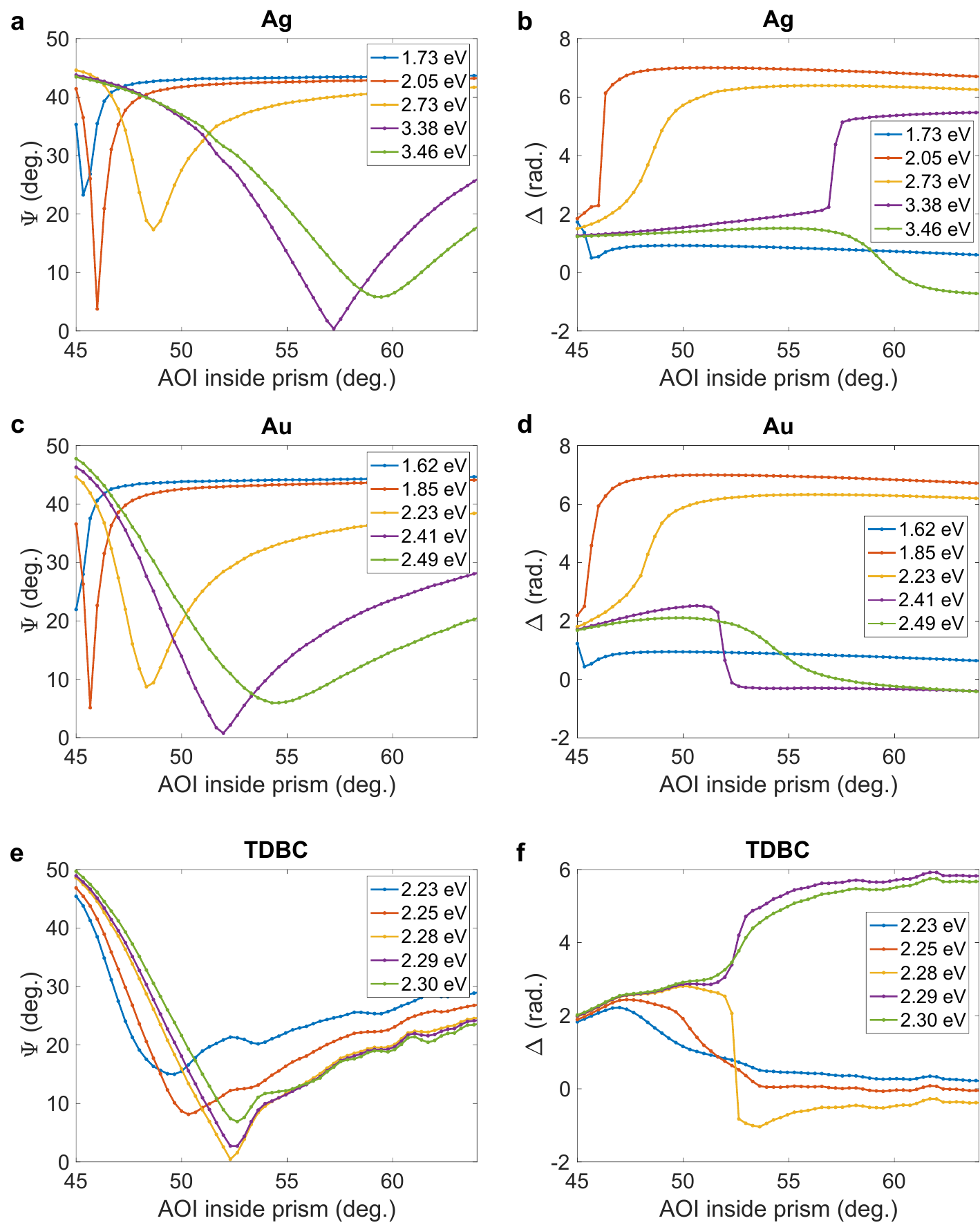}
    \caption{$\Psi$ and $\Delta$ for fixed energies as a function of incident angle.
    These plots are horizontal cross-sections of the surface plots in Figure 2 of the main manuscript.}
    \label{fig:lineaoi}
\end{figure}

In Fig. \ref{fig:lineaoi} we plot $\Psi$ (Fig. \ref{fig:lineaoi}a,c,e) and $\Delta$ (Fig. \ref{fig:lineaoi}b,d,f) as a function of incident angle.
These plots correspond to horizontal cross-sections of the plots in Fig. 2 of the main manuscript.
In the case of Ag (Fig. \ref{fig:lineaoi}a-b) and Au (Fig. \ref{fig:lineaoi}c-d), we plotted $\Psi$/$\Delta$ versus incident angle for five energies: one energy below the first phase singularity (blue lines), the energy closest to the first phase singularity (orange lines), one energy between the first and second phase singularities (yellow lines), the energy closest to the second phase singularity (purple line), and one energy higher than the second phase singularity.
In these plots, the positions of the SPPs are given by minima in $\Psi$ and jumps in $\Delta$.
As expected, the sharpest, most rapid jumps in $\Delta$ occur in plots where the minima in $\Psi$ approach zero.
The phase singularities also influence the phase response by reversing the ``direction'' of the phase jump induced by the SEPs: before the lower-energy phase singularities (blue lines in Fig. \ref{fig:lineaoi}b,d), the SPPs cause a relative drop in $\Delta$; at energies above the lower-energy phase singularity (but below the higher-energy phase singularity, yellow lines in Fig. \ref{fig:lineaoi}b,d), the SPP causes a relative increase in $\Delta$; at energies above the higher-energy phase singularity (green lines in Fig. \ref{fig:lineaoi}b,d) the SPP once again causes a decrease in $\Delta$.

\renewcommand{\thefigure}{S5}
\begin{figure}[!b]
    \centering
    \includegraphics[width=0.6\linewidth]{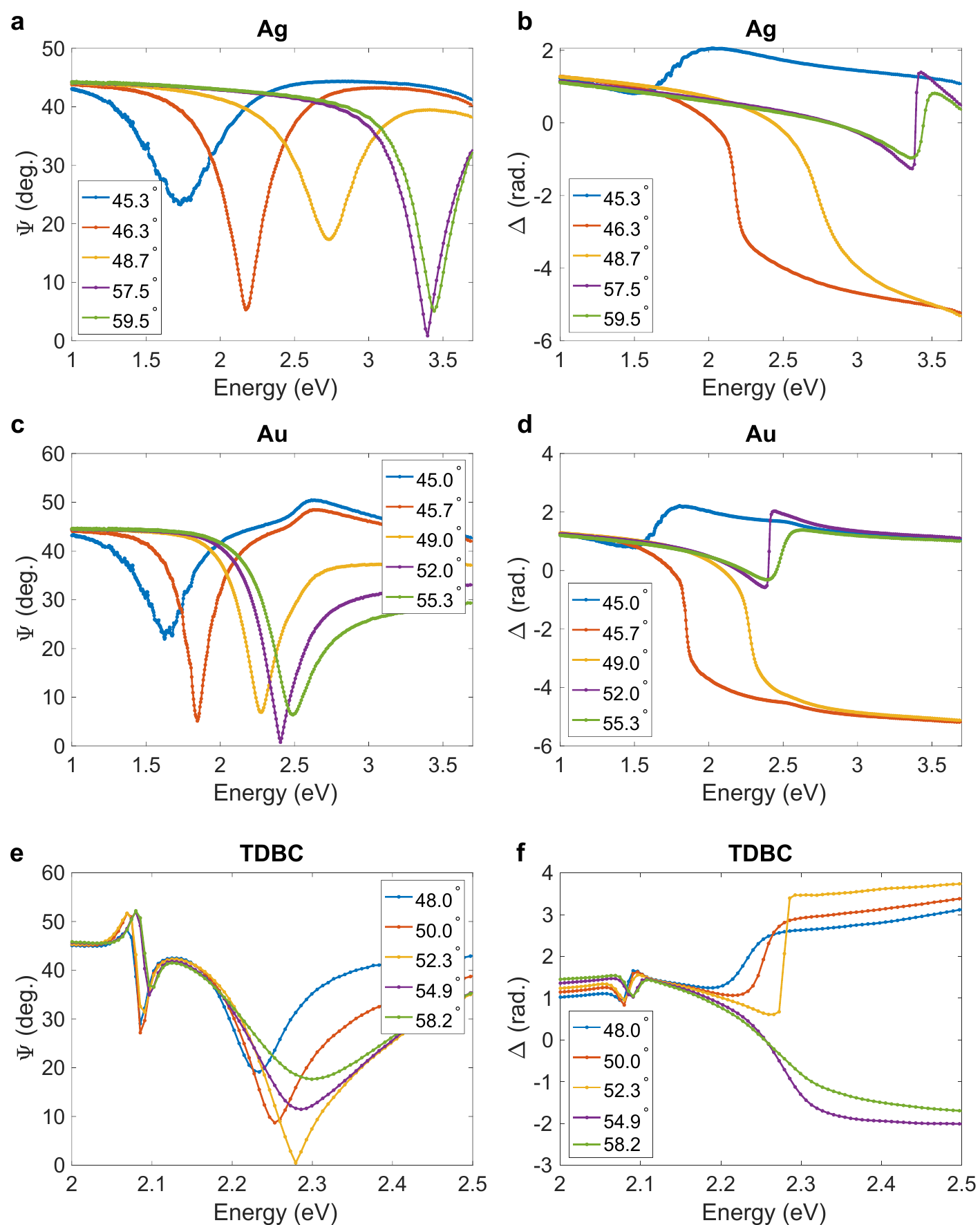}
    \caption{$\Psi$ and $\Delta$ for fixed incident angles as a function of energies.
    These plots are vertical cross-sections of the surface plots in Figure 2 of the main manuscript.}
    \label{fig:lineev}
\end{figure}

For TDBC (Fig. \ref{fig:lineaoi}e,f), we observe a similarly Fano-like response in $\Psi$; these plots are noticeable less smooth than the Ag and Au plots because of the roughness of the TDBC film.
As incident angle is increase, the illuminated area of the TDBC film changes: non-uniformity in the TDBC film thickness will therefore affect the optical response.
(The fixed-angle ellipsometry spectra plotted in Figs. \ref{fig:lineev}e,f do not suffer from this limitation and are much smoother.)
In this case, there is only one point at which the intensity drops to zero, yielding one phase singularity (yellow lines).
As with the Ag and Au SPPs, the phase singularity reverses the direction of the surface polariton-induced phase jump: $\Delta$ drops for energies below the phase singularity (blue and orange lines in Fig. \ref{fig:lineaoi}f) and increases for energies above the phase singularity (purple and green lines in in Fig. \ref{fig:lineaoi}f).

In Fig. \ref{fig:lineev}, we plot $\Psi$ and $\Delta$ spectra, i.e. we fix incident angle and plot $\Psi$ and $\Delta$ as a function of energy.
These plots correspond to vertical cross-sections of the dispersion plots in Figure 2 of the main manuscript.
As in Fig. \ref{fig:lineaoi}, for Ag and Au (Figs. \ref{fig:lineev}a-d), we plot one spectrum for an incident angle below the first phase singularity (blue lines), one spectrum close to the lower-angle phase singularity (orange lines), one spectrum between the two phase singularities (yellow lines), one spectrum close to the higher-angle phase singularity (purple lines) and one spectrum at a higher angle than the higher-angle phase singularity (green lines).
For TDBC (Figs. \ref{fig:lineev}e-f), we plot one spectrum (yellow lines) at the angle that most closely matches the phase singularity, two spectra (blue and orange lines) at angles below the phase singularity and two spectra at angles above the phase singularity (purple and green lines).
As in Fig. \ref{fig:lineaoi}, we see that phase singularities act to reverse the direction of the phase jump associated with surface polaritons.

Plotting $\Psi$ as a function of energy allows us to calculate $Q$, the quality factors for our surface modes (the ratio of a mode's energy to its spectral width)~\cite{kravets2014graphenesupp}.
The quality factor, $Q$, of the SEP mode is 14: this compares favourably with the measured quality factors for our Au ($Q\sim6$) and Ag ($Q\sim13$) SPPs.
Our observation of a phase singularity (and associated complete suppression of reflection) and relatively high $Q$ factor confirm that, in terms of quality and capacity for subwavelength field confinement, SEPs are a competitive alternative to SPPs.

\addcontentsline{toc}{subsection}{S2.3 $\rho$ spectra from prism coupling experiments}
\subsection*{S2.3 $\rho$ spectra from prism coupling experiments}

\renewcommand{\thefigure}{S6}
\begin{figure}[!h]
    %\centering
    \includegraphics[width=1.0\linewidth]{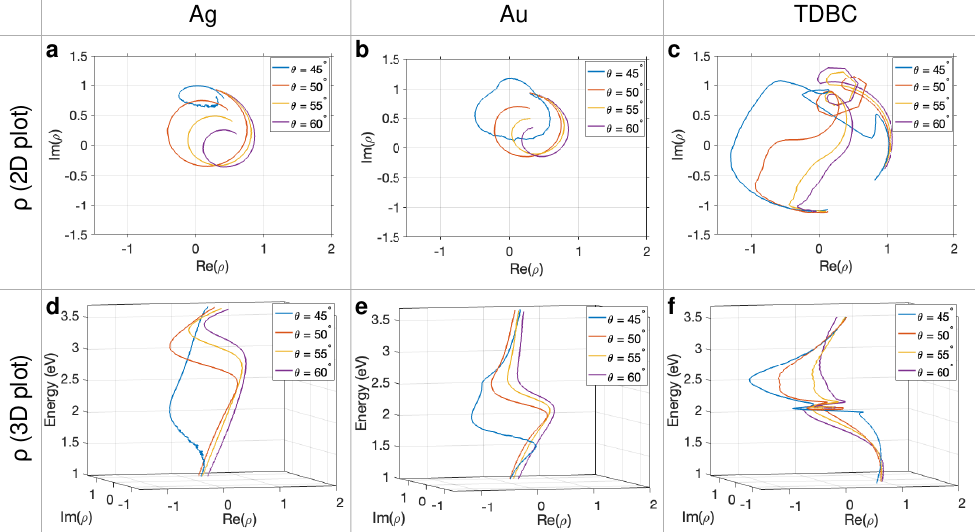}
    \caption{The combined amplitude and phase response, $\rho$, from ellipsometric prism coupling experiments in which SPPs are observed in Ag (panels a,d) and Au (panels b,e) and an SEP mode is observed in TDBC (panels c,f).
    $\rho$ is plotted with a top-down view in panels a-c (which allows for a clearer view of the topology of $\rho$~\cite{thomas2020newsupp}) and with a three-dimensional view in panels d-f (which allows one to see how $\rho$ varies with energy).}
    \label{fig:rho}
\end{figure}

\noindent
In Figure \ref{fig:rho} we plot the combined amplitude and phase response, $\rho$, for a range of incident angles in the energy range 1.0 eV $<E<$ 3.7 eV.
The first row (panels a-c) shows two-dimensional plots of $\rho$ and the second row (panels d-f) shows three-dimensional $\rho$ plots.
Surface polariton modes, which only produce a response in $r_p$, appear as an arc in $\rho$ plots.
This contrasts with Fabry-P\'{e}rot cavity modes, which give a full loop in $\rho$ because they produce a response in both $r_p$ and $r_s$~\cite{thomas2020newsupp}.
In the cases of Ag (panels a, d) and Au (panels b, e), we see two similar-looking arcs.
As incident angle is increased, in both cases the arc shifts down and then to the right.
In the two-dimensional $\rho$ plots, the arcs for Au and Ag appear very similar.
They both cross $\rho=0$ twice, resulting in two phase singularities.
The precise conditions under which $\rho=0$ (at which phase singularities are observed) can be tuned~\cite{kravets2013singularsupp}.
The $\rho$ plots for TDBC (panels d, f) are more complex.
In the top-right of panel d, we see two closed loops.
The three-dimensional $\rho$ plot in panel f makes it clear that these loops are the optical response associated with the peak in Re($\epsilon$) at 2.1 eV.
For 2.2 eV $<E<$ 2.4 eV, the spectral region in which we observed a HSEP in panel c, we observe an arc.
This arc is smaller than the arcs observed for SPPs (a consequence of the narrow band for which TDBC possesses negative real permittivity) and only crosses $\rho=0$ once because of the hyperbolic nature of HSEPs.
In their prism-coupled SPhP experiments, Carini \textit{et al.} observe a SPhP arc which crosses $\rho=0$ in a similar manner to our SEP arc, suggesting that this behaviour might be common to all narrowband surface polaritons~\cite{carini2025surfacesupp}.

\newpage
\addcontentsline{toc}{section}{S3. Neat TDBC birefringence}
\section*{S3. Neat TDBC birefringence}

\renewcommand{\thefigure}{S7}
\begin{figure}[!h]
    \centering
    \includegraphics[width=0.5\linewidth]{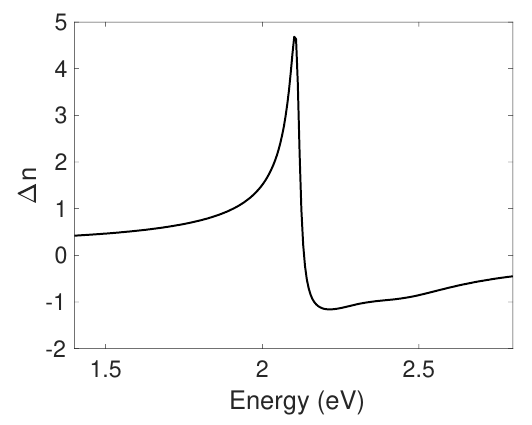}
    \caption{Birefringence of neat TDBC, determined using spectroscopic ellipsometry.}
    \label{fig:deltan}
\end{figure}

\noindent
In Fig. \ref{fig:deltan} we plot the birefringence ($\Delta n$, the difference between the in-plane and out-of-plane refractive indices) of our TDBC film.
The birefringence, $\Delta n$, of our TDBC film (plotted in Supplementary Section S3) fulfils the criterion for giant optical anisotropy ($|\Delta n| > 1$) over a wide spectral range (430 meV), with a maximum anisotropy of 4.7.
Giant optical anisotropy is regarded as a desirable property for nanophotonic materials, since it allows for the miniaturisation of optical components that rely on birefringence~\cite{niu2018giantsupp}.
The giant optical anisotropy of our TDBC films far exceeds those recently reported for some transition metal dichalcogenides~\cite{ermolaev2021giantsupp}.
These properties further highlight the potential of TDBC (and other J-aggregates) in nanophotonics beyond that of molecular strong coupling~\cite{ebbesen2016hybridsupp}.

\newpage

\newpage
\addcontentsline{toc}{section}{S4. Air/TDBC/glass ellipsometry}
\section*{S4. Air/TDBC/glass ellipsometry}

\renewcommand{\thefigure}{S8}
\begin{figure}[!h]
    \centering
    \includegraphics[width=1\linewidth]{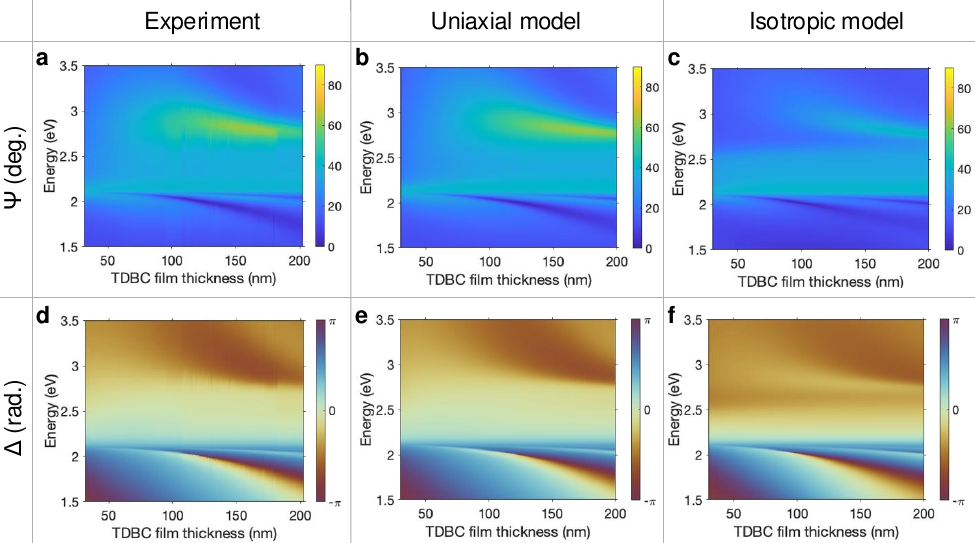}
    \caption{Ellipsometry of samples in air/TDBC/glass geometry with incident angle fixed at $65^{\circ}$ with a range of TDBC film thicknesses.
    Dispersion plots are constructed using $\Psi$ (in degrees) in panels a-c and using $\Delta$ (in radians) in panels d-f.
    Experimental results (panels a, d) are compared with transfer matrix calculations using uniaxial (panels b, e) and isotropic (panels c, f) optical constants for TDBC.}
    \label{fig:top}
\end{figure}

\noindent
To complement our prism coupling experiments in the main manuscript, here we plot the dispersion for TDBC in an air superstrate/TDBC/glass substrate configuration.
The dispersion plots in Figure \ref{fig:top} are constructed by fixing the incident angle ($65^{\circ}$) and making ellipsometric measurements of samples with a range of TDBC film thicknesses.
Each vertical strip in these dispersion plots corresponds to a film of a given thickness.

In our experimental plots (panels a, d), we observe a series of dispersive quasi-normal modes, with two TM modes (minima in $\Psi$) below 2.1 eV and one TE mode (maximum in $\Psi$) above 2.5 eV.
Our uniaxial model of TDBC (panels b, e - using the optical constants in Supplementary Section S1) very accurately reproduces all these features.
Transfer matrix calculations using our isotropic model of TDBC (panels b, f - using $\epsilon_{x,y}$ for the in-plane and out-of-plane components of the permittivity) accurately reproduces the spectral positions of the experimentally observed quasi-normal modes.
However, the quality and magnitude of the quasi-normal modes predicted by the isotropic model differ substantially from those observed in experiments.
These differences become more pronounced as the TDBC film thickness increases.
%These results provide further confirmation of the validity of our uniaxial model of TDBC.

\newpage
\addcontentsline{toc}{section}{S5. Prism-coupled hBN calculations}
\section*{S5. Prism-coupled hBN calculations}

\noindent
To explore the impact of hyperbolicity on the SPhPs supported by hBN, we performed prism coupling calculations.
For consistency with our SPP and HSEP experiments, our HSPhP calculations employed the Kretschmann configuration.
We set the refractive index for the prism in our HSPhP calculations to $n=2.37$, which approximately matches the refractive index of the Thallium bromo-iodide (KRS-5) prism used by Carini \textit{et al.} in their surface phonon polariton prism coupling experiments~\cite{carini2025surfacesupp, rodney1956refractionsupp}.
We use the optical constants for hBN determined by Caldwell \textit{et al.}~\cite{caldwell2014subsupp}.
For our anisotropic hBN calculations (main manuscript Fig. 4a-b), we used a hBN thickness of 1500 nm.
For our isotropic hBN calculations (main manuscript Fig. 4c-d), we set the out-of-plane hBN permittivity to be equal to the in-plane permittivity.
As with our TDBC calculations (main manuscript Fig. 3c-f), we needed to reduce the hBN thickness (in this case to 800 nm) to observe a SPhP in the same spectral region as the anisotropic HSPhP.

\addcontentsline{toc}{section}{S6. Plotting SP modes in the complex plane}
\section*{S6. Plotting SP modes in the complex plane}

\noindent
An alternative to plotting $\rho$ spectra (i.e. $\Psi$ and $\Delta$ spectra plotted in the complex plane, see section S2.3) is to plot the values of $\rho$ associated with each point on the SP dispersion in Fourier space.
To extract these data, we found the position of each SP mode by finding the minimum value of $\Psi$ for each incident angle.
These points trace out the SP dispersion; an example (for the HSEP mode in TDBC) is shown in Fig. \ref{fig:extractrho}.
The values of $\Psi$ that fall along this line in Fourier space are combined with their corresponding values of $\Delta$ to give $\rho$ according to $\rho = \frac{r_p}{r_s} = \tan(\Psi)e^{i\Delta}$.
The resulting curves are plotted in the complex plane in main manuscript Fig. 5.

\renewcommand{\thefigure}{S9}
\begin{figure}[!h]
    \centering
    \includegraphics[width=0.5\linewidth]{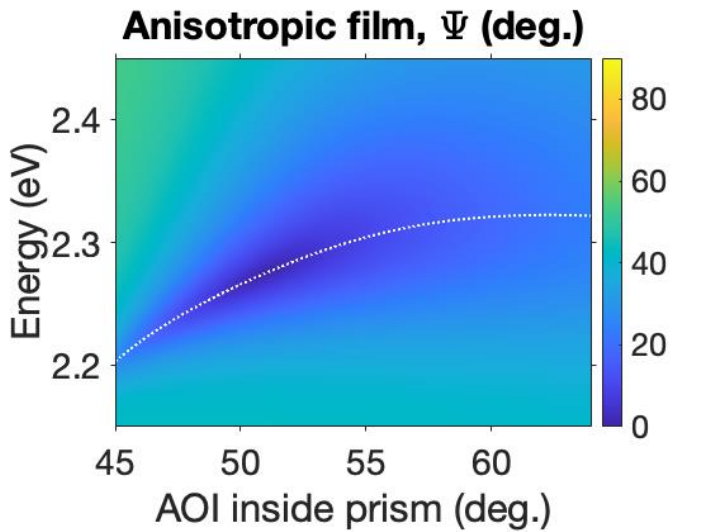}
    \caption{The calculated HSEP dispersion in $\Psi$, originally plotted in main manuscript Fig. 3c. The white dashed line is the position of the HSEP mode in Fourier space.}
    \label{fig:extractrho}
\end{figure}

\end{document}